\documentclass[aps,prl,twocolumn,superscriptaddress]{revtex4}
\usepackage{graphicx,graphics,amssymb,amsmath,hyperref}

\begin{document}

\newcommand{\s}{\hspace{-1pt}}

\def\etal#1{, #1}
\def\tit#1{#1, }
\def\jour#1{\emph{#1}}

\def\ijs{Jo\v{z}ef Stefan Institute, Jamova 39, 1000 Ljubljana, Slovenia}
\def\psi{Laboratory for Neutron Scattering and Imaging, Paul Scherrer Institut, CH-5232 Villigen, Switzerland}
\def\qmp{Department of Quantum Matter Physics, University of Geneva, CH-1211 Geneva, Switzerland}
\def\bern{Department of Chemistry and Biochemistry, University of Bern, CH-3012 Bern, Switzerland}

\title{Observation of two types of anyons in the Kitaev honeycomb magnet}

\author{N.~Jan\v{s}a}
\author{A.~Zorko}
\author{M.~Gomil\v sek}
\author{M.~Pregelj}
\affiliation{\ijs}

\author{K.~W.~Kr\"amer}
\author{D.~Biner}
\affiliation{\bern}

\author{A.~Biffin}
\affiliation{\psi}

\author{Ch.~R\"uegg}
\affiliation{\psi}
\affiliation{\qmp}

\author{M.~Klanj\v{s}ek}
\email{martin.klanjsek@ijs.si}
\affiliation{\ijs}

\date{\today}

\begin{abstract}
Quantum spin liquid is a disordered magnetic state with fractional spin excitations. Its clearest example is found in an exactly solved Kitaev honeycomb model where a spin flip fractionalizes into two types of anyons, quasiparticles that are neither fermions nor bosons: a pair of gauge fluxes and a Majorana fermion. Here we demonstrate this kind of fractionalization in the Kitaev paramagnetic state of the honeycomb magnet $\alpha$-RuCl$_3$. The spin-excitation gap measured by nuclear magnetic resonance consists of the predicted Majorana fermion contribution following the cube of the applied magnetic field, and a finite zero-field contribution matching the predicted size of the gauge-flux gap. The observed fractionalization into gapped anyons survives in a broad range of temperatures and magnetic fields despite inevitable non-Kitaev interactions between the spins, which are predicted to drive the system towards a gapless ground state. The gapped character of both anyons is crucial for their potential application in topological quantum computing.
\end{abstract}

\maketitle

\begin{figure*}
\includegraphics[width=1\linewidth]{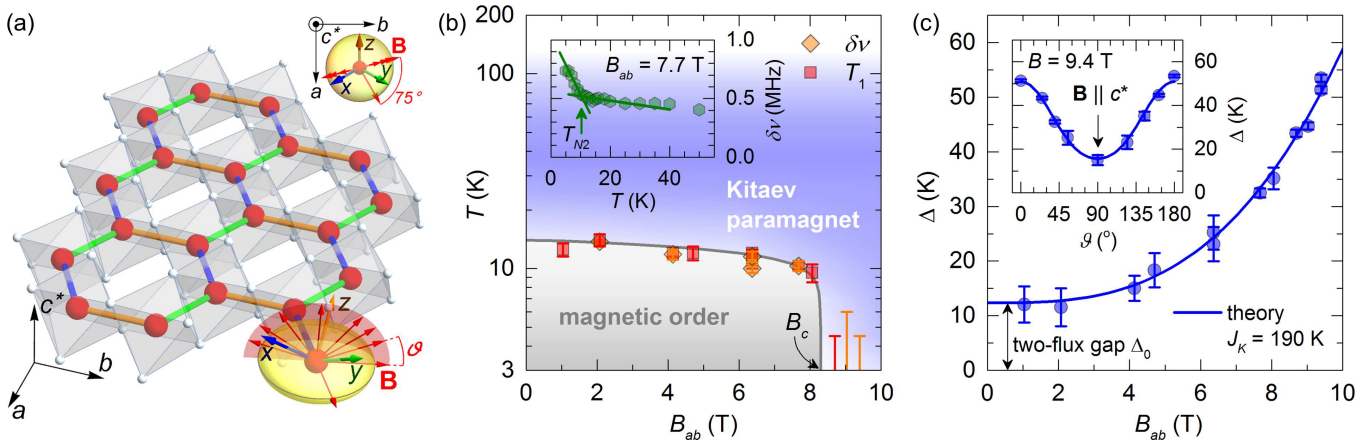}
\caption{{\bf Structure of $\alpha$-RuCl$_3$ and the key signature of anyons.} (a) The structure of a single layer of $\alpha$-RuCl$_3$ in the monoclinic $C2/m$ (no. 12) setting with the monoclinic axis $b$ ($c^*\perp a,b$). Spin-$1/2$ Ru$^{3+}$ ions (red spheres) at the centers of the edge-sharing RuCl$_6$ octahedra (gray) form an almost perfect honeycomb lattice. Ising axes of the exchange interactions between nearest-neighboring spins are perpendicular to the bond directions, pointing along $x$, $y$ or $z$ for blue, green and orange bonds, respectively. Red arrows show the employed magnetic field directions (described by the angle $\vartheta$ from the $ab$ plane) with respect to the oblate Ru$^{3+}$ $g$-tensor (yellow ellipsoid) of axial symmetry around $c^*$. The field directions form a fan (red semicircle) perpendicular to the $ab$ plane, at $15^\circ$ from the $b$ axis (inset). (b) Phase diagram of $\alpha$-RuCl$_3$ as a function of temperature $T$ and the effective magnetic field $B_{ab}=g(\vartheta)B/g_{ab}$ (so that $B_{ab}=B$ for ${\bf B}\perp c^*$) selected by the direction-dependent $g$-factor. The boundary of the magnetically ordered phase extending up to $B_c\approx 8$~T, obtained from the $^{35}$Cl linewidth $\delta\nu(T)$ (inset) and $T_1^{-1}(T)$ (Fig.~\ref{fig3}), matches the result of Ref.~\cite{Johnson_2015} (gray line). (c) The spin-excitation gap $\Delta$ as a function of $B_{ab}$ (obtained from the fits in Fig.~\ref{fig3}) follows the theoretically predicted cubic dependence (blue line) with a finite initial value corresponding to the two-flux gap $\Delta_0=0.065J_K$~\cite{Knolle_2014} with $J_K=190$~K~\cite{Do_2017}. The inset shows $\Delta(\vartheta)$ for $9.4$~T together with the curve obtained from $\Delta(B_{ab})$ (blue line). The only field direction outside the red fan in (a) is represented by the $\vartheta=180^\circ$ point.
}
\label{fig1}
\end{figure*}

In many-body systems dominated by strong fluctuations, an excitation with an integer quantum number can break up into exotic quasiparticles with fractional quantum numbers. Well known examples include fractionally charged quasiparticles in fractional quantum Hall effect~\cite{dePicciotto_1997}, spin-charge separation in one-dimensional conductors~\cite{Jompol_2009}, and magnetic monopoles in spin ice~\cite{Castelnovo_2008}. A major hunting ground for novel fractional quasiparticles are disordered magnetic states of interacting spin-$1/2$ systems governed by strong quantum fluctuations, called quantum spin liquids (QSLs). Most of their models predict that a spin-flip excitation fractionalizes into a pair of spinons, each carrying spin $1/2$~\cite{Balents_2010,Savary_2017}. Even more interesting in this respect is the Kitaev model~\cite{Kitaev_2006} of $S=1/2$ spins on a two-dimensional (2D) honeycomb lattice with nearest neighbors interacting through an Ising exchange, whose axis depends on the bond direction, as shown in Fig.~\ref{fig1}(a). This is one of a few exactly solved 2D models supporting a QSL ground state. According to the solution, a spin flip fractionalizes into a pair of bosonic gauge fluxes and a Majorana fermion~\cite{Kitaev_2006,Baskaran_2007}. As both types of quasiparticles behave as anyons, i.e., neither bosons nor fermions, under exchange, they could be used for decoherence-free topological quantum computation~\cite{Kitaev_2006}. The experimental detection of such anyons is thus the primary goal of current QSL research.

As fractional quasiparticles are always created in groups, their common signature is a continuous spin-excitation spectrum, observed in recent QSL candidates on the kagome and triangular lattices~\cite{Han_2012,Paddison_2016}, instead of sharp magnon modes found in ordered magnets. A Kitaev QSL also exhibits this feature~\cite{Knolle_2014,Nasu_2016}, as well as additional, specific signatures, all related to the fact that fractionalization in this case leads to different quasiparticles. First, the fractionalization proceeds in two steps, with both types of quasiparticles releasing their entropy at different temperatures~\cite{Nasu_2015}. Second, although Majorana fermions themselves are gapless in zero magnetic field, the response of the QSL to a spin flip is gapped due to the inevitable simultaneous creation of a pair of gapped gauge fluxes~\cite{Knolle_2014}. And third, in the presence of an external magnetic field, the Majorana fermions also acquire a gap, which is predicted to grow with the characteristic third power of the field in the low-field region~\cite{Kitaev_2006,Jiang_2011,Nasu_2017}. Currently, $\alpha$-RuCl$_3$ stands as the most promising candidate for the realization of the Kitaev QSL~\cite{Nasu_2016,Sandilands_2015,Banerjee_2016,Banerjee_2017,Do_2017}. Among the listed signatures, a spin-excitation continuum was observed by Raman spectroscopy~\cite{Sandilands_2015,Nasu_2016} and inelastic neutron scattering~\cite{Banerjee_2016,Banerjee_2017,Do_2017}, and the two-step thermal fractionalization was confirmed by specific-heat measurements~\cite{Do_2017}, all in zero field. However, an application of a finite field, which should affect the gaps of both types of quasiparticles differently, is crucial to identify them. Using nuclear magnetic resonance (NMR), we determine the field dependence of the spin-excitation gap $\Delta$ shown in Fig.~\ref{fig1}(c), which indeed exhibits a finite zero-field value predicted for gauge fluxes and the cubic growth predicted for Majorana fermions. This result clearly demonstrates the fractionalization of a spin flip into two types of anyons in $\alpha$-RuCl$_3$.

$\alpha$-RuCl$_3$ is structurally related to the other two Kitaev QSL candidates, Na$_2$IrO$_3$~\cite{Choi_2012} and $\alpha$-Li$_2$IrO$_3$~\cite{Singh_2012}. All three are layered Mott insulators based on the edge-sharing octahedral units, RuCl$_6$ and IrO$_6$ [Fig.~\ref{fig1}(a)], respectively, and driven by strong spin-orbit coupling~\cite{Plumb_2014}, which together lead to a dominant Kitaev exchange coupling between the effective $S=1/2$ spins of Ru$^{3+}$ and Ir$^{4+}$ ions, respectively~\cite{Jackeli_2009}. A monoclinic distortion of the IrO$_6$ octahedra in both iridate compounds results in the presence of non-Kitaev exchange interactions between the spins, which lead to the low-temperature magnetic ordering and thus prevent the realization of the QSL ground state. Judging by the lower transition temperature, these interactions are smaller in $\alpha$-RuCl$_3$~\cite{Kubota_2015,Sears_2015,Majumder_2015,Cao_2016}. Signatures of fractional quasiparticles should thus be sought in a region of the phase diagram outside the magnetically ordered phase, at temperatures where the Kitaev physics is not yet destroyed by thermal fluctuations. This is the Kitaev paramagnetic region [Fig.~\ref{fig1}(b)] extending to a relatively high temperature around $100$~K where the nearest-neighbor spin correlations vanish~\cite{Do_2017}.

\begin{figure}
\includegraphics[width=1\linewidth]{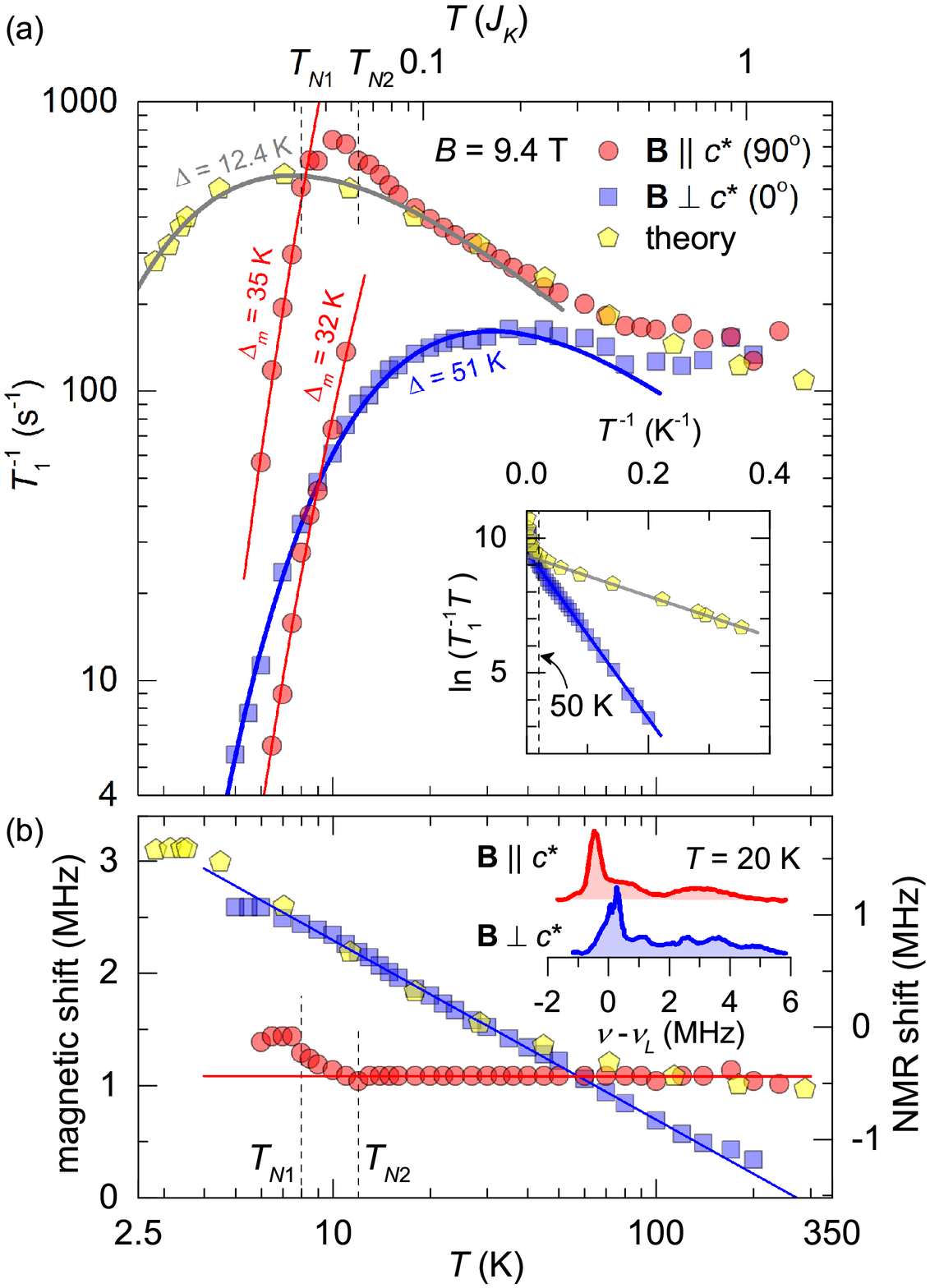}
\caption{{\bf Evidence for the spin-excitation continuum.} (a) $^{35}$Cl $T_1^{-1}$ as a function of temperature $T$ in $9.4$~T for two magnetic field orientations, ${\bf B}\parallel c^*$ ($B_{ab}=4.1$~T) and ${\bf B}\perp c^*$ ($B_{ab}=9.4$~T). The theoretical prediction~\cite{Yoshitake_2016} using $J_K=190$~K~\cite{Do_2017} is rescaled in vertical direction to match the ${\bf B}\parallel c^*$ dataset between $17$~K and $100$~K. Dashed lines mark the transition temperatures $T_{N2}=12$~K ($AB$ stacking) and $T_{N1}=8$~K ($ABC$ stacking) into the magnetically ordered states. Red lines are fits to $T_1^{-1}\propto T^2\exp(-\Delta_m/T)$ for gapped magnon excitations in the 3D ordered state. Blue and gray lines are fits to Eq.~(\ref{mod}) for fractional spin excitations in the Kitaev paramagnet valid up to $50$~K. Inset demonstrates the resulting linear dependence of ${\ln}(T_1^{-1}T)$ on $T^{-1}$ below $50$~K. (b) Temperature dependent $^{35}$Cl magnetic shift (i.e., NMR shift with subtracted quadrupole shift~\cite{Supplemental}) of the dominant NMR peak (the inset shows the whole central line) in $9.4$~T for two field orientations compared to the theoretical prediction~\cite{Yoshitake_2016} using $J_K=190$~K~\cite{Do_2017} and rescaled in vertical direction to match the ${\bf B}\perp c^*$ dataset between $6$~K and $50$~K. Red and blue lines are phenomenological linear fits in the semi-log scale.
}
\label{fig2}
\end{figure}

The boundary of the magnetically ordered phase measured in a large $\alpha$-RuCl$_3$ single crystal~\cite{Supplemental} using $^{35}$Cl NMR is displayed in Fig.~\ref{fig1}(b). Magnetic properties of $\alpha$-RuCl$_3$ are known to be highly anisotropic~\cite{Johnson_2015,Majumder_2015}, mainly because of the anisotropic Ru$^{3+}$ $g$-tensor [Fig.~\ref{fig1}(a)] with $g_{ab}=2.5$ and $g_{c^*}=1.1$~\cite{Yadav_2016}. We exploit this anisotropy to scan the phase diagram by varying the direction of the applied fixed field, instead of varying the magnitude of the field applied in the $ab$ plane~\cite{Johnson_2015}. Namely, as the Zeeman term contains the product $gB$, a magnetic field $B$ applied at an angle $\vartheta$ from the $ab$ plane is equivalent to the effective field $B_{ab}=g(\vartheta)B/g_{ab}$ applied in the $ab$ plane, where $g(\vartheta)=\sqrt{g_{ab}^2\cos^2{\vartheta}+g_{c^*}^2\sin^2{\vartheta}}$ is the direction-dependent $g$-factor. This is valid if the studied underlying physics is close to isotropic, a condition to be verified at the end. As shown in the inset of Fig.~\ref{fig1}(b), we determine the transition temperature $T_{N2}$ as the onset of NMR line broadening~\cite{Supplemental} monitored on the dominant NMR peak [inset of Fig.~\ref{fig2}(b)]. The obtained phase boundary extending up to the critical field $B_c\approx 8$~T matches the result of a recent reference study~\cite{Johnson_2015}. The observed transition temperature $T_{N2}$ of around $14$~K near zero field is consistent with a considerable presence of the two-layer $AB$ stacking in the monoclinic $C2/m$ crystal structure [Fig.~\ref{fig1}(a)], in addition to the three-layer $ABC$ stacking, which is characterized by a lower transition temperature $T_{N1}$ of around $7$~K in zero field~\cite{Banerjee_2016,Cao_2016}. As our study is focused on the Kitaev paramagnetic region [Fig.~\ref{fig1}(b)] governed by the physics of individual layers, it is not affected by the particular stacking type.

To detect and monitor the spin-excitation gap as a function of the magnetic field, we use the NMR spin-lattice relaxation rate $T_1^{-1}$, which directly probes the low-energy limit of the local spin-spin correlation function and thus offers a direct access to the spin-excitation gap. Fig.~\ref{fig2}(a) shows the $^{35}$Cl $T_1^{-1}(T)$ datasets taken on the dominant NMR peak [inset of Fig.~\ref{fig2}(b)] in $9.4$~T for two magnetic field orientations. A noticeable feature of the $T_1^{-1}(T)$ dataset for ${\bf B}\perp c^*$ (i.e., in the $ab$ plane, $B_{ab}=9.4$~T) is a broad maximum around $30$~K, followed by a steep decrease towards lower temperatures. In the $T_1^{-1}(T)$ dataset for ${\bf B}\parallel c^*$, such a feature would apparently develop at a lower temperature, if the dataset was not disrupted by the phase transition at $T_{N2}=12$~K [in a field of $B_{ab}=4.1$~T, Fig.~\ref{fig2}(b)]. Instead, two $T_1^{-1}$ components develop below $T_{N2}$, both exhibiting a steep drop, one below $T_{N2}$ and the other one below $T_{N1}=8$~K. These two phase transitions were observed before and ascribed to the presence of $AB$ and $ABC$ stackings, respectively~\cite{Banerjee_2016,Cao_2016}. The analysis of the data below $T_{N2}$ and $T_{N1}$ using the expression $T_1^{-1}\propto T^2\exp(-\Delta_m/T)$ valid for gapped magnon excitations in the 3D ordered state~\cite{Supplemental} gives comparable values of the magnon gap $\Delta_m=32$~K and $35$~K, respectively, implying the same low-energy physics in both cases. The obtained values are compatible with the gap of $29$~K determined by inelastic neutron scattering~\cite{Banerjee_2016,Ran_2017}.

\begin{figure*}
\includegraphics[width=1\linewidth]{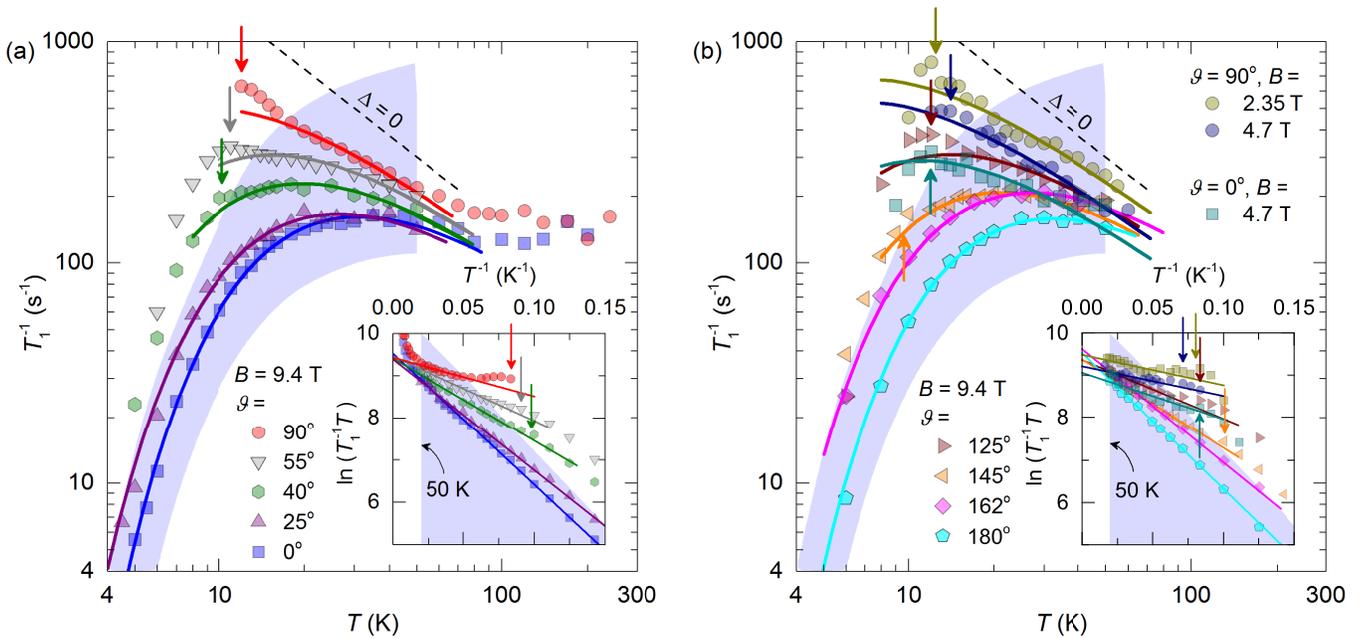}
\caption{{\bf Determination of the spin-excitation gap $\Delta$.} (a,b) $^{35}$Cl $T_1^{-1}$ as a function of temperature $T$ in $2.35$, $4.7$ and $9.4$~T for various magnetic field orientations given by $\vartheta$. Arrows mark the transition temperatures $T_{N2}$ into the magnetically ordered state, defined by a weakly pronounced onset of a $T_1^{-1}$ decrease on decreasing $T$. Solid lines are fits to Eq.~(\ref{mod}) for fractional spin excitations in the Kitaev paramagnet, between the temperature slightly above $T_{N2}$ and $50$~K (blue background). These allow us to determine $\Delta(\vartheta)$ and $\Delta(B_{ab})$ dependencies shown in Fig.~\ref{fig1}(c). Dashed line is the curve $T_1^{-1}\propto T^{-1}$ defined by $\Delta=0$ showing the largest negative slope. Insets demonstrate the resulting linear dependence of ${\ln}(T_1^{-1}T)$ on $T^{-1}$ in the appropriate $T^{-1}$ range (blue background). The $\vartheta=180^\circ$ dataset corresponds to the only field direction outside the red fan in Fig.~\ref{fig1}(a).
}
\label{fig3}
\end{figure*}

To access the key information held by the $T_1^{-1}(T)$ datasets in the Kitaev paramagnetic state, we first observe in Fig.~\ref{fig2}(a) that the dataset for ${\bf B}\perp c^*$ below $100$~K exhibits the same shape as the theoretical dataset numerically calculated for the ferromagnetic Kitaev model in zero field~\cite{Supplemental,Yoshitake_2016}. A characteristic broad maximum of the latter is a sign of thermally excited pairs of gauge fluxes over the two-flux gap~\cite{Yoshitake_2016}, whose exact value amounts to $\Delta_0=0.065J_K$~\cite{Kitaev_2006,Knolle_2014} where $J_K$ is the Kitaev exchange coupling. As shown in Fig.~\ref{fig2}(a), a large part of the theoretical dataset, up to around $0.2J_K$, well above the maximum, can be excellently described by the phenomenological expression
\begin{equation}\label{mod}
	T_1^{-1}\propto\frac{1}{T}\exp\left(-\frac{n\Delta}{T}\right),
\end{equation}
where $n$ is set to $0.61$ in order for $\Delta$ to match the required value of $\Delta_0$. This expression has a useful property: its maximum appears at a temperature $\Delta/n$, which allows for a simple estimate of $\Delta$ directly from the $T_1^{-1}(T)$ dataset. The ${\bf B}\perp c^*$ ($B_{ab}=9.4$~T) dataset up to $50$~K is excellently reproduced by Eq.~(\ref{mod}) using $\Delta=51$~K [Fig.~\ref{fig2}(a)]. The validity of Eq.~(\ref{mod}) in this case and in the case of the theoretical dataset is clearly demonstrated in the inset of Fig.~\ref{fig2}(a), which shows the resulting linear dependence of ${\ln}(T_1^{-1}T)$ on $T^{-1}$ below $50$~K. Meanwhile, even in the absence of the characteristic maximum, the ${\bf B}\parallel c^*$ ($B_{ab}=4.1$~T) dataset above $17$~K, i.e., slightly above $T_{N2}$, and up to high temperatures matches the theoretical zero-field dataset using the value $J_K=190$~K determined by inelastic neutron scattering, also based on the ferromagnetic Kitaev model~\cite{Do_2017}. This means that the gap for $4.1$~T is already close to the zero-field value $\Delta_0=0.065J_K=12.4$~K. A large difference between the two determined gaps points to a significant $\Delta(B_{ab})$ variation in the Kitaev paramagnetic state. Finally, the temperature-independent part of both $T_1^{-1}(T)$ datasets above $100$~K indicates a crossover into the classical paramagnetic state~\cite{Moriya_1956}, in line with the result of Ref.~\cite{Do_2017}.

The expression given by Eq.~(\ref{mod}) is not merely phenomenological, but reveals the presence of gapped fractional spin excitations. Similar expressions are obtained for the $T_1$ relaxation due to gapped magnons in magnetic insulators at low temperatures $T\ll\Delta$~\cite{Supplemental}. In this case, the prefactor $T^{-1}$ is replaced by a more general $T^p$ originating from the magnon density of states $g(E)$, which depends on the dimensionality $D$, while $n$ is generally the number of magnons involved in the process. For $n=1$ (single-magnon scattering) and a quadratic dispersion relation for magnons, one obtains $g(E)\propto E^{D/2-1}$ and thus $p=D-1\geq 0$, while higher $n$ (multi-magnon scattering) lead to even higher powers $p$~\cite{Supplemental}. At higher temperatures $T\sim\Delta$, the effective $p$ changes, but always remains positive. As the very unusual $p=-1$ in Eq.~(\ref{mod}) valid for $T\lesssim\Delta$ cannot be obtained for magnons, fractional spin excitations should be involved. This is furthermore supported by a fractional $n$ in Eq.~(\ref{mod}), implying that fractions of a spin-flip excitation are involved in the relaxation process. In contrast to this unusual gapped $T_1^{-1}(T)$ behavior, the temperature dependence of the local susceptibility monitored by the $^{35}$Cl NMR shift in Fig.~\ref{fig2}(b) is monotonic over the whole covered temperature range, as predicted for the ferromagnetic Kitaev model~\cite{Yoshitake_2016}. Such a dichotomy between the two observables is a direct sign of spin fractionalization, as different fractional quasiparticles enter the two observables in different ways~\cite{Yoshitake_2016}.

To obtain the spin-excitation gap $\Delta$ as a function of $B_{ab}$ in Fig.~\ref{fig1}(c), the $T_1^{-1}(T)$ datasets in Fig.~\ref{fig3} taken in magnetic fields of different directions and magnitudes are fitted to Eq.~(\ref{mod}) in the temperature range of the Kitaev paramagnetic phase. As the curve $T_1^{-1}\propto T^{-1}$ defined by $\Delta=0$ is steeper than any dataset in this range, the obtained excitation gaps are apparently all nonzero. The inset of Fig.~\ref{fig1}(c) showing the symmetric $\Delta(\vartheta)$ dependence around $90^\circ$ in $9.4$~T, where $\vartheta$ traverses nonequivalent directions with respect to the Kitaev axes on both sides [inset of Fig.~\ref{fig1}(a)], demonstrates that the underlying physics is indeed isotropic as assumed when introducing $B_{ab}$. The obtained $\Delta(B_{ab})$ in Fig.~\ref{fig1}(c) can be perfectly reproduced as a sum of two terms: the two-flux gap $\Delta_0$ and the gap acquired by Majorana fermions in a weak magnetic field, predicted to be proportional to the cube of the field~\cite{Kitaev_2006,Jiang_2011,Nasu_2017,Supplemental},
\begin{equation}\label{cube}
  \Delta=\Delta_0+\frac{\alpha}{3}\frac{\widetilde{B}^3}{\Delta_0^2},
\end{equation}
using $J_K=190$~K~\cite{Do_2017} to evaluate $\Delta_0$, as before, while $\widetilde{B}=g_{ab}\mu_BB_{ab}/k_B$ is the field in kelvin units, $k_B$ is the Boltzmann constant, $\mu_B$ the Bohr magneton, and $\alpha=4.5$ (leading to the best fit) accounts for the sum over the excited states in the third-order perturbation theory, which is the origin of the $\widetilde{B}^3$ term~\cite{Kitaev_2006}. This result demonstrates that a spin-flip excitation in $\alpha$-RuCl$_3$ indeed fractionalizes into a gauge-flux pair and a Majorana fermion.

Focusing on the Kitaev paramagnetic region in the phase diagram of $\alpha$-RuCl$_3$ in Fig.~\ref{fig1}(b) is essential for our observation of two types of anyons. Instead, other recent experimental studies focused on the low-temperature region above $B_c$, observing the spin-excitation continuum~\cite{Banerjee_2017_2} with either a gapless behavior~\cite{Leahy_2017,Zheng_2017} or the gap opening linearly with $B-B_c$~\cite{Baek_2017,Sears_2017,Hentrich_2017,Ponomaryov_2017}, but without a definite conclusion about the identity of the involved quasiparticles. Such an ambiguous behavior likely originates in the presence of additional, smaller non-Kitaev interactions between the spins~\cite{Banerjee_2016,Yadav_2016,Winter_2016,Ran_2017}, whose role should be pronounced particularly at low temperatures and which are indeed predicted to drive the system towards a gapless QSL ground state~\cite{Song_2016}. Our result shows that spin fractionalization into two types of anyons is robust against these interactions in a broad range of temperatures and magnetic fields. This is the main practical advantage of $\alpha$-RuCl$_3$ with respect to all other anyon realizations, such as the fractional quantum Hall effect in 2D heterostructures~\cite{dePicciotto_1997} or hybrid nanowire devices~\cite{Mourik_2012}, where anyons are observed only at extremely low temperatures and at certain field values. Our discovery thus establishes $\alpha$-RuCl$_3$ as a unique platform for future investigations of anyons and braiding operations on them, which form the functional basis of a topological quantum computer~\cite{Kitaev_2006}.

\begin{acknowledgments}
The work was partly supported by the Slovenian ARRS program No. P1-0125 and project No. PR-07587.
\end{acknowledgments}

\newpage\clearpage
\section{Supplemental material}

\subsection{Crystal growth}

Crystals of $\alpha$-RuCl$_3$ were synthesized from anhydrous RuCl$_3$ (Strem Chemicals). The starting material was heated in vacuum to $200$~$^\circ$C for one day to remove volatile impurities. In the next step, the powder was sealed in a silica ampoule under vacuum and heated to $650$~$^\circ$C in a tubular furnace. The tip of the ampoule was kept at lower temperature and the material sublimed to the colder end during one week. Phase pure $\alpha$-RuCl$_3$ (with a high-temperature phase of $C2/m$ crystal structure) was obtained as thin crystalline plates. The residual in the hot part of the ampoule was black RuO$_2$ powder. The purified $\alpha$-RuCl$_3$ was sublimed for the second time in order to obtain bigger crystal plates. The phase and purity of the compounds was verified by powder X-ray diffraction. All handling of the material was done under strictly anhydrous and oxygen-free conditions in glove boxes or sealed ampoules. Special care has to be taken when the material is heated in sealed-off ampoules. If gas evolves from the material, this may result in the explosion of the ampoule.

\subsection{Magnetic susceptibility}

\begin{figure}
\includegraphics[width=1\linewidth]{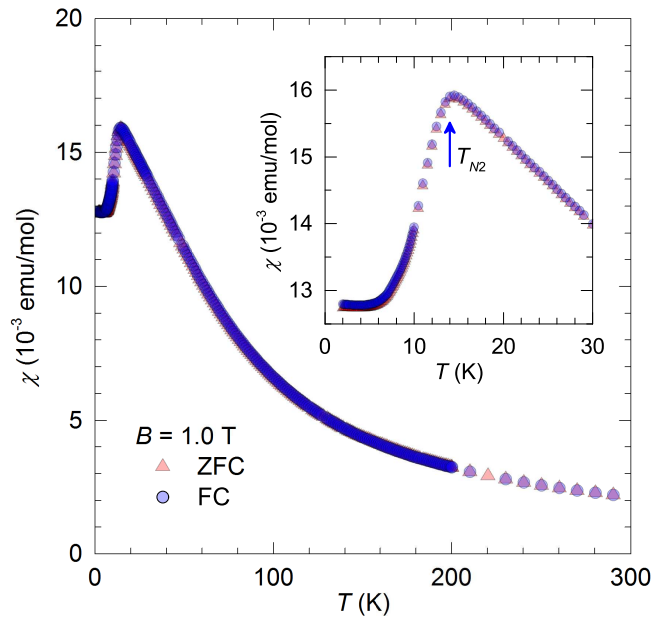}
\caption{{\bf Magnetic susceptibility.} Zero-field-cooled (ZFC) and field-cooled (FC) magnetic susceptibility of the powdered $\alpha$-RuCl$_3$ sample in a magnetic field of $1.0$~T. Inset is a zoom around the magnetic phase transition at $T_{N2}=14$~K.
}
\label{figS1}
\end{figure}

Magnetic susceptibility measurements of $\alpha$-RuCl$_3$ were performed using a Quantum Design MPMS. A powdered sample of the mass $22.5$~mg was placed into a plastic capsule, in a glovebox to avoid contact with air, and then quickly transferred into the MPMS. Fig.~\ref{figS1} shows the measured susceptibility taken with cooling in field and in zero field. The obtained curve with the magnetic transition at $T_{N2}=14$~K (inset of Fig.~\ref{figS1}) is almost identical to the corresponding curve in Ref.~\cite{Johnson_2015}.

\subsection{Nuclear magnetic resonance}

{\bf General.} The $^{35}$Cl nuclear magnetic resonance (NMR) experiments were performed on a foil-like $\alpha$-RuCl$_3$ single crystal of approximate dimensions $5\times 5\times 0.1$~mm$^3$ in a continuous-flow cryostat allowing us to reach temperatures down to $4.2$~K. When handling the sample, we took extreme care to minimize its exposure to air. A thin NMR coil fitting the sample size was made from a thin copper wire with $20$-$40$ turns, depending on the required tuning frequency determined by the external magnetic field. The coil was covered with a mixture of epoxy and ZrO$_2$ powder, which was allowed to harden, in order to ensure the rigidity of the coil. The coil was then mounted on a teflon holder attached to a rotator, which allowed us to vary the orientation of the sample with respect to the external magnetic field. In order to reduce the noise of an already weak $^{35}$Cl NMR signal, a consequence of the extremely broad $^{35}$Cl NMR spectrum, we used a bottom-tuning scheme. With the output radio-frequency power of around $20$~W, the typical $\pi/2$ pulse length amounted to $2~\mu$s. The NMR signals were recorded using the standard spin-echo, $\pi/2-\tau_d-\pi$ pulse sequence with a typical delay of $\tau_d=70~\mu$s (much shorter than the spin-spin relaxation time $T_2$) between the $\pi/2$ and $\pi$ pulses.

{\bf $T_1$ relaxation.} The spin-lattice relaxation (i.e., $T_1$) experiment was carried out using an inversion recovery pulse sequence, $\varphi_i-\tau-\pi/2-\tau_d-\pi$, with an inversion pulse $\varphi_i<\pi$ (suitable for broad NMR lines) and a variable delay $\tau$ before the read-out spin-echo sequence. The spin-lattice relaxation datasets were typically taken at $20$ increasing values of $\tau$. The datasets were analyzed using the model of magnetic relaxation for $I=3/2$ spin monitored on the central $-1/2\longleftrightarrow 1/2$ transition:
\begin{equation}\label{T1}
	m(\tau)=1-(1+s)\left[0.1\exp\left(-\frac{\tau}{T_1}\right)+0.9\exp\left(-\frac{6\tau}{T_1}\right)\right],
\end{equation}
where $T_1$ is the spin-lattice relaxation time and $s$ is the inversion factor. In the region of the phase diagram outside the magnetically ordered phase [Fig.~2(b)], this expression reproduces the experimental relaxation curves perfectly. In the magnetically ordered phase, two $T_1$ components appear, and the relaxation curves are reproduced as a sum of two terms of the form given by Eq.~(\ref{T1}). For instance, the temperature dependence of the corresponding two $T_1$ values for $B=9.4$~T with ${\bf B}\parallel c^*$ is given in Fig.~2(a). In cases where only a narrow temperature region below the transition was covered, the two components in the relaxation curves were hard to identify, and we used Eq.~(\ref{T1}) with a stretched exponent instead.

\begin{figure}
\includegraphics[width=1\linewidth]{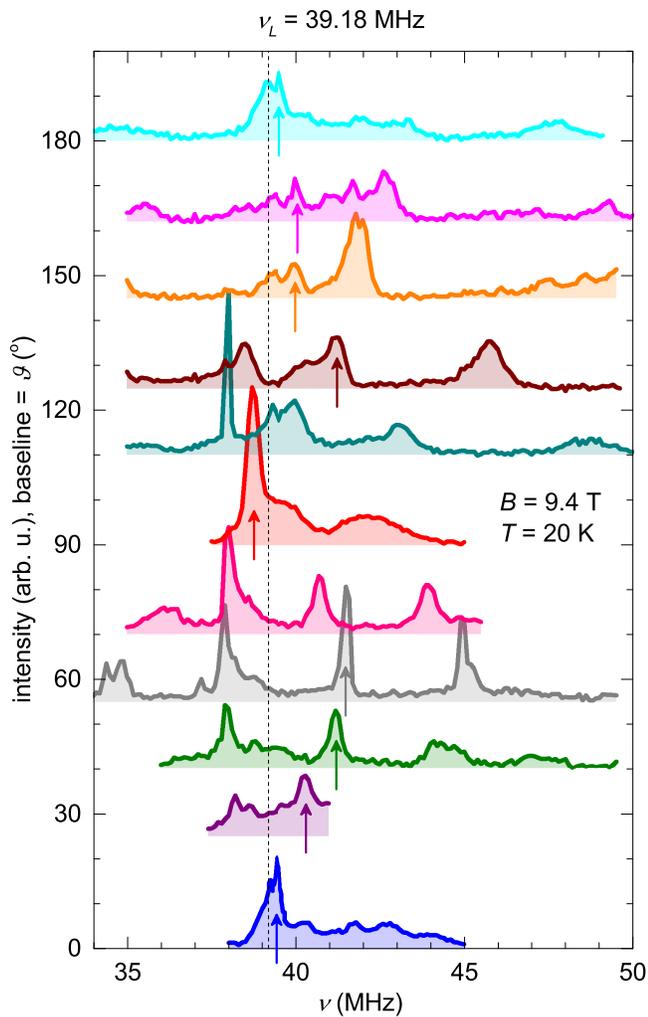}
\caption{{\bf Orientation dependence of the NMR spectrum.} Central line of the $^{35}$Cl NMR spectrum of the $\alpha$-RuCl$_3$ single crystal taken at $20$~K in the field of $9.4$~T applied at an angle $\vartheta$ with respect to the crystal $ab$ plane [as in Fig.~1(a)]. The plane of rotation is at an angle of $15^\circ$ from the crystal $b$ axis [as shown in Fig.~1(a)]. Dashed vertical line indicates the Larmor frequency $\nu_L$. Arrows mark the peak whose temperature dependence is analyzed in Fig.~\ref{figS3} and where $T_1$ displayed in Fig.~3 was measured. The $\vartheta=180^\circ$ spectrum corresponds to the only field direction outside the red fan in Fig.~1(a).
}
\label{figS2}
\end{figure}

{\bf Orientation dependence of the NMR spectrum.} The $^{35}$Cl NMR spectra were recorded point by point in frequency steps of $50$ or $100$~kHz, so that the Fourier transform of the signal was integrated at each step to arrive at the individual spectral point. The covered NMR frequency range was from $34$~MHz, the lower limit of our setup, up to $50$~MHz. The dependence of the corresponding part of the NMR spectrum on the direction of the external magnetic field of $9.4$~T (described by the angle $\vartheta$ from the crystal $ab$ plane) at a temperature of $20$~K is shown in Fig.~\ref{figS2}. The spectra are extremely broad because of large $^{35}$Cl (with $I=3/2$ spin) quadrupole interaction. As concluded in the following, a large portion of the covered frequency range is associated with the central, $1/2\longleftrightarrow -1/2$ $^{35}$Cl NMR transition. As this transition is observed to consist of at least three peaks (Fig.~\ref{figS2}), even for the symmetric ${\bf B}\parallel c^*$ orientation (with $\vartheta=90^\circ$), while there are only two inequivalent Cl sites in the crystal structure, the splitting of the central line is likely a consequence of stacking faults in the layered crystal structure or crystal twinning, or both.

\begin{figure*}
\includegraphics[width=1\linewidth]{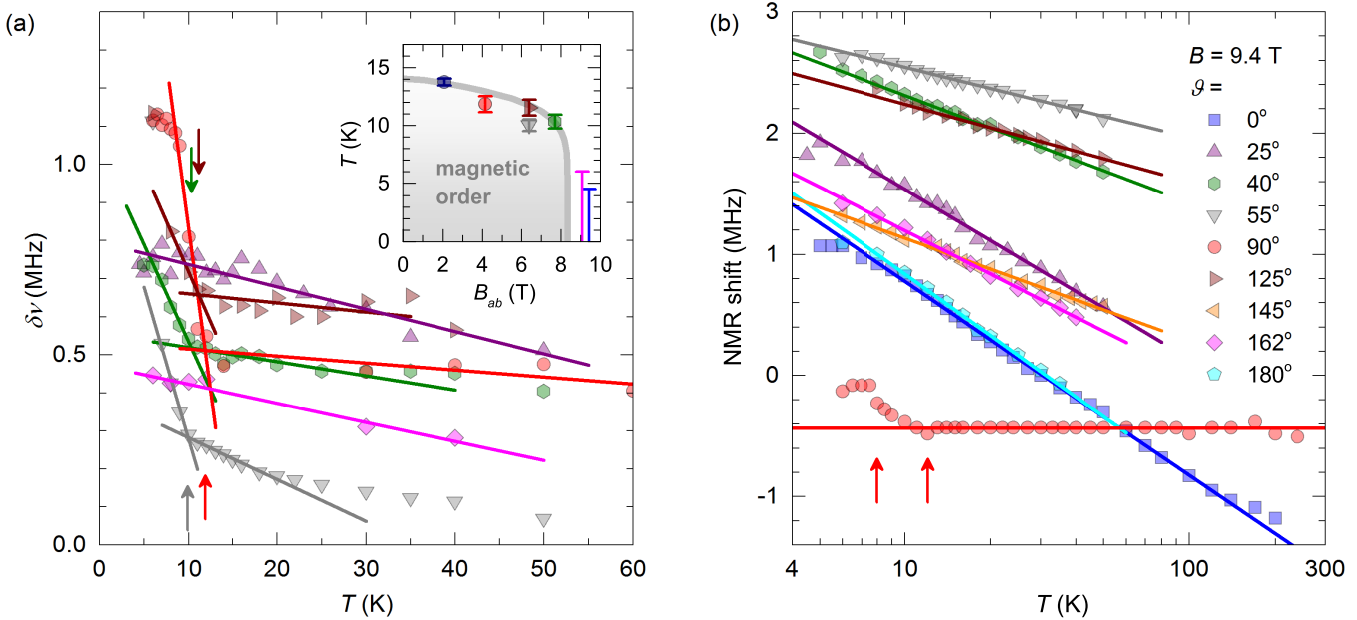}
\caption{{\bf Temperature dependence of the NMR line.} The temperature ($T$) dependence of (a) the width $\delta\nu$ and (b) the NMR frequency shift of dominant $^{35}$Cl NMR peaks (marked by arrows in Fig.~\ref{figS2}) in the field of $9.4$~T ($^{35}$Cl Larmor frequency $\nu_L=39.18$~MHz) for various sample orientations given by $\vartheta$. The overlaping lines for some values of $\vartheta$ do not allow the determination of $\delta\nu$. For each dataset drawn in (a), straight lines are linear fits on both sides of the kink at $T_{N2}$ (determined as the temperature of intersection and marked by an arrow) indicating the onset of low-temperature magnetic ordering. Inset shows the obtained points of the phase boundary [same symbols as the corresponding $\delta\nu(T)$ datasets] compared to the result of Ref.~\cite{Johnson_2015} (gray line). Straight lines in (b) are phenomenological linear fits. Only the $\vartheta=90^\circ$ (${\bf B}\parallel c^*$) dataset in (b) shows signs of magnetic transitions (marked by arrows).
}
\label{figS3}
\end{figure*}

{\bf Relation bewteen orientation and field dependence of $T_1$.} Measuring the $T_1$ dependence on the direction of the magnetic field (described by the angle $\vartheta$ from the crystal $ab$ plane) instead of on its magnitude in the $ab$ plane allows us to cover low $B_{ab}$ values, while keeping the applied magnetic field $B$ high. This is beneficial for two reasons related to the strong quadrupole broadening of the $^{35}$Cl NMR spectrum (Fig.~\ref{figS2}): to minimize an already large NMR linewidth and to keep the Larmor frequency well above the quadrupole splitting, which is of the order of $10$~MHz as concluded in the following. The validity of this approach is supported by the fact that $\Delta(B_{ab})$ data points for various angles $\vartheta$ and field values $2.35$, $4.7$ and $9.4$~T in Fig.~1(c) all collapse on a smooth experimental curve. The $\Delta(B_{ab})$ data points taken in lower fields apparently exhibit much larger error bars. Namely, the corresponding $T_1^{-1}(T)$ datasets in Fig.~3(b) are more scattered than the datasets taken in $9.4$~T despite a much longer averaging for noise reduction.

{\bf Temperature dependence of the NMR line.} We measured the temperature dependence of the dominant $^{35}$Cl NMR peak in a field of $9.4$~T for various sample orientations. From these measurements, we determined the temperature dependence of the frequency width [Fig.~\ref{figS3}(a)] and the NMR shift of the peak with respect to the Larmor frequency [Fig.~\ref{figS3}(b)]. For $B_{ab}<8$~T, the width exhibits a clear kink as a function of temperature, which indicates the onset of NMR line broadening at the phase transition into the magnetically ordered state. Plotting the temperature of the kink as a function of $B_{ab}$ in the inset of Fig.~\ref{figS3}, we obtain the phase boundary of the magnetically ordered state, which perfectly matches the result of the reference study~\cite{Johnson_2015}. In contrast, the NMR shift does not exhibit any signs of a magnetic transition, except for the $\vartheta=90^\circ$ (${\bf B}\parallel c^*$) dataset. We find the NMR shift to be a monotonic function of temperature $T$, empirically following a ${\rm log}\,T$ dependence over a broad temperature range.

\begin{figure}
\includegraphics[width=1\linewidth]{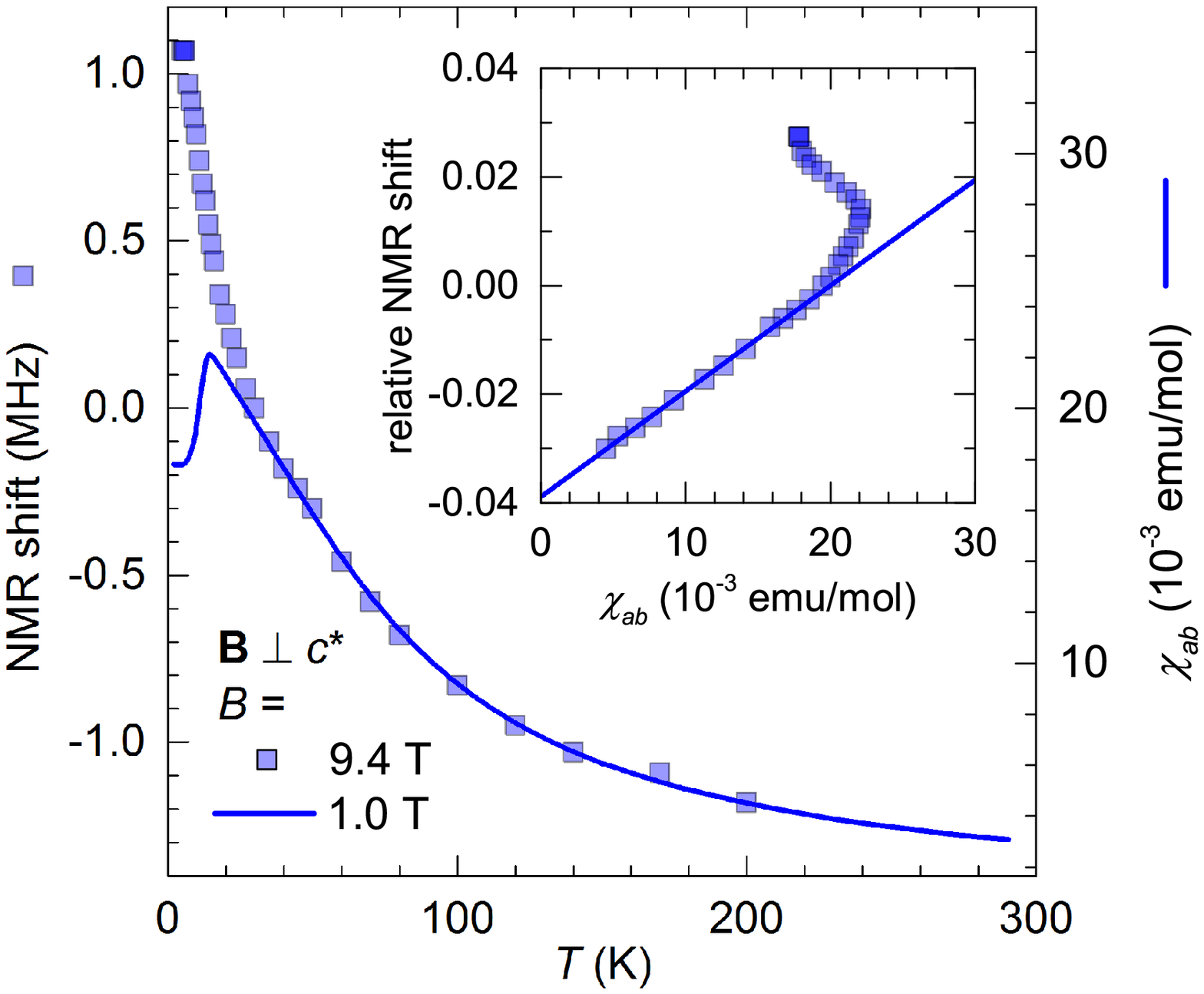}
\caption{{\bf NMR shift against susceptibility.} Temperature dependences of the $^{35}$Cl NMR shift and magnetic susceptibility $\chi_{ab}$ are proportional to each other down to $35$~K. Inset shows the dependence of the relative NMR shift (i.e., NMR shift divided by $\nu_L=39.18$~MHz) on $\chi_{ab}$. Line is a linear fit of the dataset up to $20\cdot 10^{-3}$~emu/mol, i.e., down to $35$~K.
}
\label{figS4}
\end{figure}

{\bf Contributions to the NMR shift.} To separate the magnetic contribution to the NMR frequency shift from the temperature-independent quadrupole contribution, we plot the relative NMR shift (i.e., the NMR shift divided by the $^{35}$Cl Larmor frequency $\nu_L=39.18$~MHz in a field of $9.4$~T) measured on the dominant NMR peak for ${\bf B}\perp c^*$, i.e., in the $ab$ plane [the $\vartheta=0^\circ$ dataset in Fig.~\ref{figS3}(b)], against the rescaled magnetic susceptibility $\chi_{ab}$ in the inset of Fig.~\ref{figS4}. In Ref.~\cite{Johnson_2015}, an experimental ratio between the susceptibility $\chi$ of the powdered sample and the susceptibility $\chi_{ab}$ of the single crystal with a field applied in the $ab$ plane is obtained as $(2+r)/3$ with $r=0.157$, leading to $\chi_{ab}=3\chi/(2+r)$. We use this empirical relation to evaluate $\chi_{ab}(T)$ from our field-cooled $\chi(T)$ dataset shown in Fig.~\ref{figS1}. As we did not measure susceptibility in high magnetic fields, we rely on the dataset taken in $1.0$~T. This is valid in a broad temperature range, except at low temperatures where this dataset starts to deviate from the high-field susceptibility~\cite{Kubota_2015}. From the observed linear relation between the relative shift and $\chi_{ab}$ up to $20\cdot 10^{-3}$~emu/mol (i.e., down to $35$~K), we obtain the hyperfine coupling constant $A=2.2$~T/$\mu_B$ and the zero-temperature relative shift $-0.039$ that, when multiplied by $\nu_L$, gives the quadrupole shift $\Delta\nu_Q=-1.53$~MHz.

From the obtained quadrupole shift $\Delta\nu_Q$, we can estimate the quadrupole splitting $\nu_Q$ between the successive $^{35}$Cl NMR transitions. For the case of an axially symmetric EFG tensor and the field applied at an angle $\vartheta'$ from the principal EFG axis $v_{ZZ}$ with the largest EFG eigenvalue, the second-order quadrupole shift is given by $\Delta\nu_Q=-3\nu_Q^2(1-\cos\vartheta'^2)(9\cos\vartheta'^2-1)/(16\nu_L)$ for the $I=3/2$ nucleus~\cite{Abragam_2011}. As the axes of the EFG tensor are not known, we assume a typical tilt $45^\circ$ of $v_{ZZ}$ from $c^*$, so that $\vartheta'\sim 45^\circ$. From the previously evaluated $\Delta\nu_Q$ we then obtain $\nu_Q\sim 14.2$~MHz. This is an estimate of the quadrupole splitting between the central $^{35}$Cl NMR transition and the satellite transitions. We can thus conclude that the NMR peaks in the covered frequency range of Fig.~\ref{figS2} all belong to the central transition.

\subsection{Theory}

{\bf Theoretical $T_1^{-1}(T)$ curve.} The theoretical temperature dependence of $T_1^{-1}$ is numerically calculated for the Kitaev model in zero field~\cite{Yoshitake_2016}. $T_1^{-1}$ contains two contributions, one coming from a single fluctuating spin (i.e., on-site) and the other one coming from fluctuating nearest-neighboring (NN) spins in the Kitaev honeycomb lattice. As the $^{35}$Cl nucleus in $\alpha$-RuCl$_3$ is located at equal distances from the closest two Ru$^{3+}$ $S=1/2$ spins [Fig.~1(a)], $T_1^{-1}$ contains both contributions with equal weights. Namely, as the hyperfine coupling constant $A$ of $^{35}$Cl to both spins is the same, the relevant spin-spin correlation function can be generally written as
\begin{align}\label{corr}
	\Bigl< A&\bigl\{S_1(t)\pm S_2(t)\bigr\}\cdot A(S_1\pm S_2)\Bigr> = \nonumber\\
	&= A^2\Bigl[ \bigl< S_1(t)S_1\bigr> + \bigl< S_2(t)S_2\bigr> \,\pm \nonumber\\
	&\pm \Bigl(\bigl< S_1(t)S_2\bigr> + \bigl< S_2(t)S_1\bigr>\Bigr) \Bigr],
\end{align}
for the involved components $S_1$ and $S_2$ of both Ru$^{3+}$ spins, where the plus (minus) sign is valid for ferromagnetic (antiferromagnetic) fluctuations. The first two terms on the right side of Eq.~(\ref{corr}) represent the on-site contributions, while the last two represent the NN-sites contributions, both with apparently equal weights. Accordingly, the theoretical curve for the ferromagnetic case, plotted in Fig.~2(a), is the average of the on-site and NN-sites contributions.

{\bf $T_1$ relaxation due to gapped magnons.} When spin fluctuations in the magnetic lattice are due to excited magnons, the corresponding spin-lattice relaxation rate for a single-magnon process is given by~\cite{Beeman_1968}
\begin{equation}\label{beeman}
	T_1^{-1}\propto\int g^2(E)n(E)\bigl[1+n(E)\bigr]{\rm d}E,
\end{equation}
where $E$ is the energy of magnons, $g(E)$ is their density of states, $n(E)=[\exp(\beta E)-1]^{-1}$ is the Bose-Einstein distribution function, $\beta=1/(k_B T)$, and $k_B$ is the Boltzmann constant. Denoting the magnon gap by $\Delta$ (in kelvin units), we define $\varepsilon=E-k_B\Delta$ as the energy measured from the bottom of the magnon band. The power-law dispersion relation $\varepsilon\propto k^s$ in $D$ dimensions, which includes the standard parabolic dispersion ($s=2$) and the Dirac dispersion ($s=1$) as special cases, leads to $g(E)\propto\varepsilon^{D/s-1}$. For low temperatures $T\ll\Delta$, the distribution function $n(E)$ can be approximated by the Boltzmann distribution, $n(E)\approx\exp(-\beta E)=\exp(-\Delta/T)\exp(-\beta\varepsilon)$. Plugging these expressions for $g(E)$ and $n(E)$ into Eq.~(\ref{beeman}), we obtain
\begin{equation}\label{simpleT1}
	T_1^{-1}\propto T^{2D/s-1}\exp\left(-\frac{\Delta}{T}\right)\int_0^\infty\exp(-x)\,x^{2(D/s-1)}\,{\rm d}x.
\end{equation}
The integral on the right side of Eq.~(\ref{simpleT1}) converges if $s<2D$ and evaluates to $\Gamma(2D/s-1)$ where $\Gamma$ is the gamma function. We can thus rewrite Eq.~(\ref{simpleT1}) as
\begin{equation}\label{simpleT1p}
	T_1^{-1}\propto T^p\exp\left(-\frac{\Delta}{T}\right)
\end{equation}
with the power of the prefactor $p=2D/s-1$. In case of $D=2$, which is relevant for the Kitaev honeycomb magnet, $p=1$ for $s=2$ and $p=3$ for $s=1$, so that the power $p$ cannot be negative. Even in case of $D=1$, $p$ can only reach the lowest value of $0$ precisely for $s=2$ [although care should be taken in this case, as the integral in Eq.~(\ref{simpleT1}) then formally diverges]. If more than a single magnon is involved in the $T_1$ process, the power $p$ is also positive and becomes even higher~\cite{Beeman_1968}. Gapped magnons thus cannot lead to the $T_1$ relaxation described by Eq.~(\ref{simpleT1p}) with $p<0$.

Instead, we can use Eq.~(\ref{simpleT1p}) in the 3D magnetically ordered state, when the elementary excitations are magnons with a gap $\Delta_m$. In this case $D=3$ and $s=2$, and this leads to $T_1^{-1}\propto T^2\exp(-\Delta_m/T)$. We use this expression to analyze the $T_1^{-1}(T)$ data [Fig.~2(a)] in the low-temperature ordered state of $\alpha$-RuCl$_3$.

As a side observation, all these examples show that a frequently used simple gapped model $T_1^{-1}\propto\exp(-\Delta_s/T)$ with the gap $\Delta_s$, which was used before to analyze the $T_1^{-1}(T)$ datasets in $\alpha$-RuCl$_3$~\cite{Baek_2017}, is actually not justified in any region of the phase diagram of $\alpha$-RuCl$_3$.

\begin{figure}
\includegraphics[width=1\linewidth]{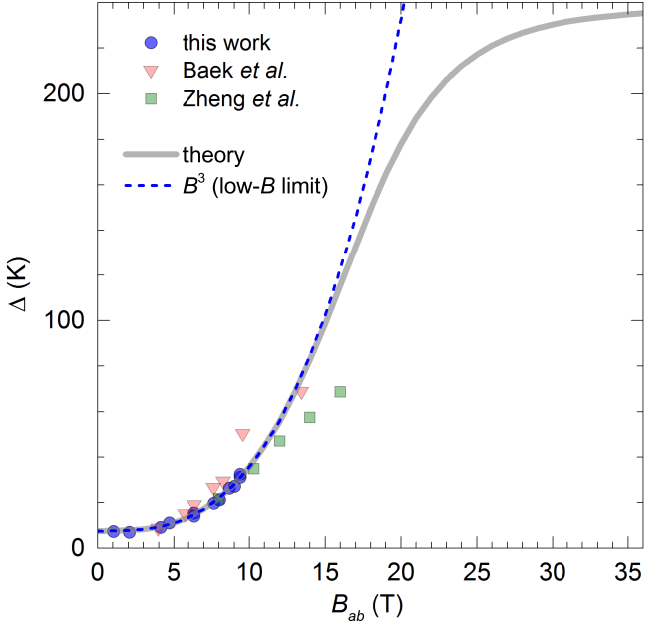}
\caption{{\bf Spin-excitation gap.} The spin-excitation gap $\Delta$ as a function of the magnetic field $B$ applied in the crystal $ab$ plane obtained from $T_1^{-1}(T)$ data in our work and in two recent works~\cite{Baek_2017,Zheng_2017} using our model given by Eq.~(1). The data are compared to the theoretical expression (using $J_K=190$~K~\cite{Do_2017}, $\alpha=4.5$, $g=g_{ab}$), which simplifies to the $B^3$ dependence given by Eq.~(2) [i.e., $\Delta_0$ added to $\Delta_f$ in Eq.~(\ref{cubeF})] in the low-field region.
}
\label{figS5}
\end{figure}

{\bf Majorana fermion gap.} In the Kitaev model, Majorana fermions acquire a gap in the presence of an external magnetic field~\cite{Kitaev_2006}. This is shown for a field applied perpendicularly to the honeycomb plane, i.e., in the $(1,1,1)$ direction in the coordinate system defined by the Kitaev axes $x$, $y$ and $z$. The corresponding Zeeman term then reads $\mathcal{H}_Z=-h\sum_j (S_j^x+S_j^y+S_j^z)$, where $h=g\mu_BB/\sqrt{3}$ is a single component of the magnetic field $B$ in energy units, $g$ is the $g$-factor and $\mu_B$ is the Bohr magneton. When treated as a perturbation to the Kitaev Hamiltonian, the Zeeman term contributes to the Majorana fermion gap only at third order~\cite{Kitaev_2006}. The corresponding effective Hamiltonian is thus proportional to $h^3$ and can be written as~\cite{Kitaev_2006,Jiang_2011,Nasu_2017}
\begin{equation}\label{third}
	\mathcal{H}_{\rm eff}^{(3)}=-\alpha\frac{h^3}{k_B^2\Delta_0^2}\sum_{jkl}S_j^xS_k^yS_l^z,
\end{equation}
where $\Delta_0$ is a two-flux gap (in kelvin units), while $\alpha$ (of the order of unity) accounts for the sum over the excited states, and its exact value is not known. The Kitaev model extended with such a three-spin exchange term $-\kappa\sum_{jkl}S_j^xS_k^yS_l^z$ with $\kappa=\alpha h^3/(k_B^2\Delta_0^2)$ is still exactly solvable and the dispersion relation of the Majorana fermions is calculated as~\cite{Jiang_2011}
\begin{equation}\label{dispersion}
	E_{\bf k}=2\sqrt{k_B^2J_K^2\vert 1+e^{i{\bf k}\cdot{\bf a}_1}+e^{i{\bf k}\cdot{\bf a}_2}\vert^2
	+\kappa^2\sin^2({\bf k}\cdot{\bf a}_1)},
\end{equation}
where $J_K$ is the Kitaev coupling (in kelvin units), while ${\bf a}_1$ and ${\bf a}_2$ are the unit vectors of the honeycomb lattice. The dispersion relation given by Eq.~(\ref{dispersion}) is gapped for $\kappa\neq 0$, and the corresponding gap $\Delta_f$ can be calculated numerically as a function of $\kappa$ and thus as a function of the magnetic field. For small magnetic fields, i.e., for $\kappa\ll k_BJ_K$, the Majorana fermion gap (in kelvin units) simplifies to
\begin{equation}\label{cubeF}
	\Delta_f=\sqrt{3}\frac{\kappa}{k_B}=\frac{\alpha}{3\Delta_0^2}\left(\frac{g\mu_BB}{k_B}\right)^3,
\end{equation}
while for high magnetic fields it saturates to $\Delta_f=2J_K$. The total spin-excitation gap $\Delta$ is obtained by adding $\Delta_f$ to the two-flux gap $\Delta_0$. The field dependence of $\Delta$ is shown in Fig.~\ref{figS5} for $J_K=190$~K (taken from Ref.~\cite{Do_2017} and used in this work), $g=g_{ab}$ and $\alpha=4.5$ [leading to the best fit of our $\Delta(B_{ab})$ data points]. The cubic approximation given by Eq.~(\ref{cubeF}), which is also plotted in Fig.~\ref{figS5}, is apparently valid up to $15$~T, well beyond the field range covered in this work.

\subsection{Comparison with recent works}

{\bf Recent NMR works.} Very recently, two $^{35}$Cl NMR studies of $\alpha$-RuCl$_3$ appeared~\cite{Baek_2017,Zheng_2017}. The analysis of both studies is focused on the low-temperature region below $15$~K. Using a simple exponential model $T_1^{-1}\propto\exp(-\Delta_s/T)$ in this region, Ref.~\cite{Baek_2017} finds that the excitation gap $\Delta_s$ opens linearly with the field above the critical field around $10$~T. On the other hand, Ref.~\cite{Zheng_2017} extends the covered temperature range down to $1.5$~K and finds a low-temperature gapless, power-law behavior of $T_1^{-1}(T)$ in the covered high-field region above the critical field around $8$~T.

As these results are very different from our results, also because of quite different analysis, we analyze the $T_1^{-1}(T)$ datasets obtained in these two works also with our model given by Eq.~(1). As in our work, we focus on the Kitaev paramagnetic region, to the temperature range from $50$~K down to slightly above the transition temperature below $8$~T, and down to $4.2$~K above $8$~T (or a bit higher at higher fields), including the characteristic maximum in $T_1^{-1}(T)$ as a main feature. The data in Ref.~\cite{Zheng_2017} were taken with a field applied in the crystal $ab$ plane, while the data in Ref.~\cite{Baek_2017} were taken with a field applied at $30^\circ$ and $-60^\circ$ with respect to the $ab$ plane. In this case, we calculate the effective field values $B_{ab}$ in the same way as in our work. The obtained field dependence of the excitation gap $\Delta(B_{ab})$ for both works is shown in Fig.~\ref{figS5} together with our result. Applying our analysis to the data in all three works apparently leads to relatively consistent results. Nevertheless, the results for the data from Refs.~\cite{Baek_2017,Zheng_2017} alone do not allow to conclude on the cubic field dependence of $\Delta$, mostly due to the lack of important low-field data points. Regarding Ref.~\cite{Baek_2017}, the two data points at the highest fields seem to deviate from the trend set by the other points. The corresponding two $T_1^{-1}(T)$ datasets exhibit a suspicious plateau at low temperatures, not observed in any other dataset of the three works, which casts some doubts on their credibility. Regarding Ref.~\cite{Zheng_2017}, the obtained $\Delta(B_{ab})$ data points, which complement our field region, nicely continue the trend set by our data points. The obtained trend is approximately linear in field, consistent with the theoretical prediction in this intermediate-field region, although with a different slope. However, the theoretical prediction is based on a perturbative treatment, which may not give reliable results outside the low-field region.

In light of our conclusions, the reported findings of these two works should be understood in the following way. While the true signs of fractionalization into Majorana fermions and gauge fluxes can be found in the Kitaev paramagnetic region covering a broad temperature range up to around $100$~K [Fig.~1(a)], the physics at low temperatures of the order of $10$~K and below is apparently obscured, most likely due to the effect of inevitable non-Kitaev interactions, as predicted already in Ref.~\cite{Zheng_2017}.

\clearpage

\end{document}